# Galactic Charge

S. Reucroft[1]
ThinkIncubate, Inc., Wellesley, Mass., USA

September, 2014.  Revised and expanded, April, 2015.

Abstract

We investigate the hypothesis that the core of a galaxy has a positive electrical charge with an equal and opposite negative charge distributed over the galactic periphery.  We present a determination of the amount of charge needed to explain the apparent anomalous rotation behaviour.

--------------

It is often stated that galactic rotation curves provide strong evidence for the existence of dark matter.  Figure 1 shows a typical galactic rotation curve (solid line, labeled Meas).  It is based on measurements on our own Milky Way galaxy [1].  Other galactic rotation curves are very similar [2].  All of these curves show an initial increase in rotation speed as the distance from galactic center is increased with the rotation speed being approximately constant outside the galactic core.  Taking into account the visible distribution of galactic matter, these curves are not consistent with the Kepler Law.  For example, the dotted line in figure 1 (labeled Kepler) shows a theoretical rotation curve that is based on the Milky Way visible galactic mass distribution shown in figure 2.

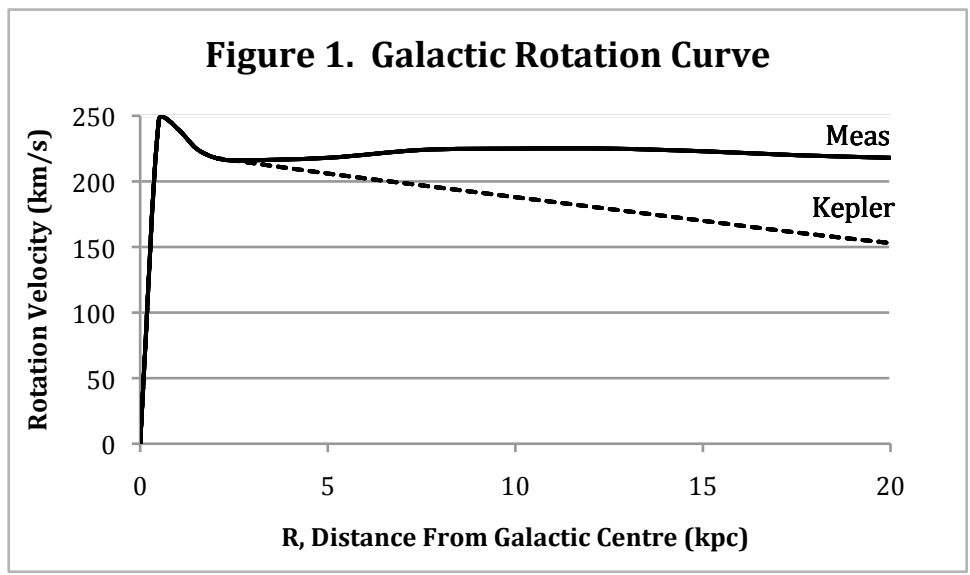

The Kepler rotation curve shows a reduction in rotation velocity outside the galactic core and this is not consistent with data.  This has led to the suggestion that most of the galactic matter is composed of some hitherto unknown substance that interacts gravitationally but not electromagnetically. This is usually referred to as dark matter.

---

[1] reucroft@gmail.com

We have investigated the possibility that these galaxies exhibiting anomalous rotation behavior are, in fact, electrically non-isotropic and that the apparent anomalous behavior stems from attractive electrostatic forces between core and periphery.

We have developed a simple galactic model that incorporates a separation of charge between core and periphery. The model consists of a series of masses $m_i$ with charge $q_i$ in orbit at distances $R_i$ from the galactic core of mass $M$ and charge $Q$. Figure 2 shows the galactic mass distribution that we have used for our model. This is the mass distribution used to make the curve labeled Kepler in figure 1.

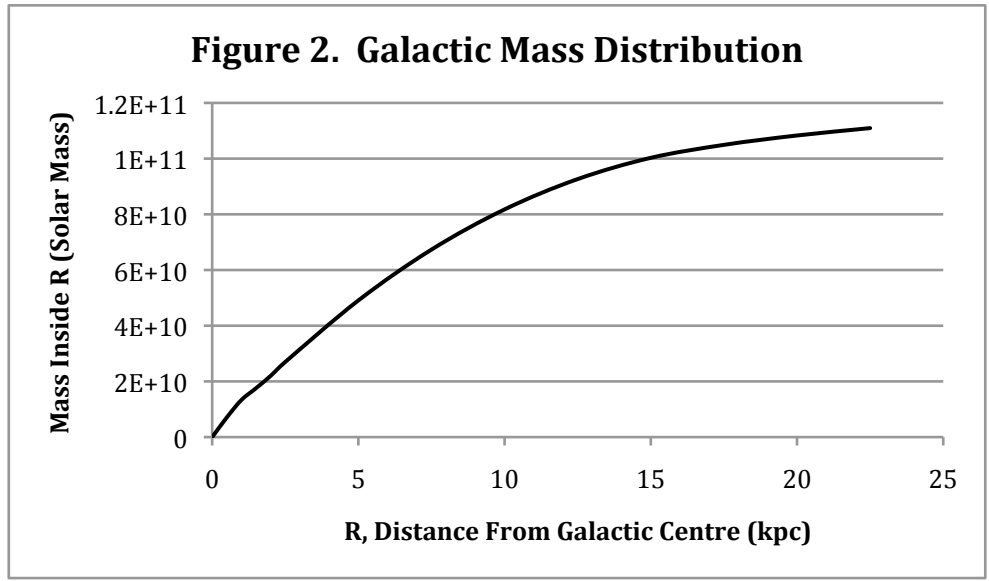

The galactic equation of motion is given by:

$$\frac{m_i v_{1i}^2}{R_i} = \frac{GMm_i}{R_i^2} + \frac{kQq_i}{R_i^2},$$

which gives:

$$v_{1i}^2 = \frac{GM}{R_i} + \frac{kQq_i}{m_i R_i},$$

where $v_{1i}$ is the orbital velocity, $G$ is the Newton gravitation constant and $k$ the Coulomb electrostatic constant ($=1/4\pi\varepsilon_0$).

Our hypothesis is that $v_{1i}$ is the observed galactic rotation velocity at a distance $R_i$ from the galactic centre.

In the presence of gravity alone, the equation of motion would be:

$$\frac{m_i v_{2i}^2}{R_i} = \frac{GMm_i}{R_i^2},$$

which gives:

$$v_{2i}^2 = \frac{GM}{R_i}.$$

This is equivalent to Kepler's Law and $v_{2i}$ is the Keplerian rotation velocity.

Replacing $GM/R_i$ in the expression for $v_{1i}^2$ with $v_{2i}^2$ and rearranging, we obtain:

$$Qq_i = \frac{m_i R_i}{k}(v_{1i}^2 - v_{2i}^2).$$

We use an iterative process to determine the $q_i$ using the $m_i$ from figure 2 and the rotation velocities from figure 1.

The first step is to determine an initial approximation for $Q$. This is accomplished by using a 2-body approximation that the core holds a charge $+Q$, and the rest of the galaxy has charge $-Q$ and orbits at a distance $R$ from the galactic centre, where $R$ is midway between centre and edge. We obtain $Q \sim 10^{31}$ C. The $q_i$ are then obtained, the final values of which are shown in figure 3. An important cross-check is given by summing the $q_i$ and verifying that $Q = \sum_i q_i$.

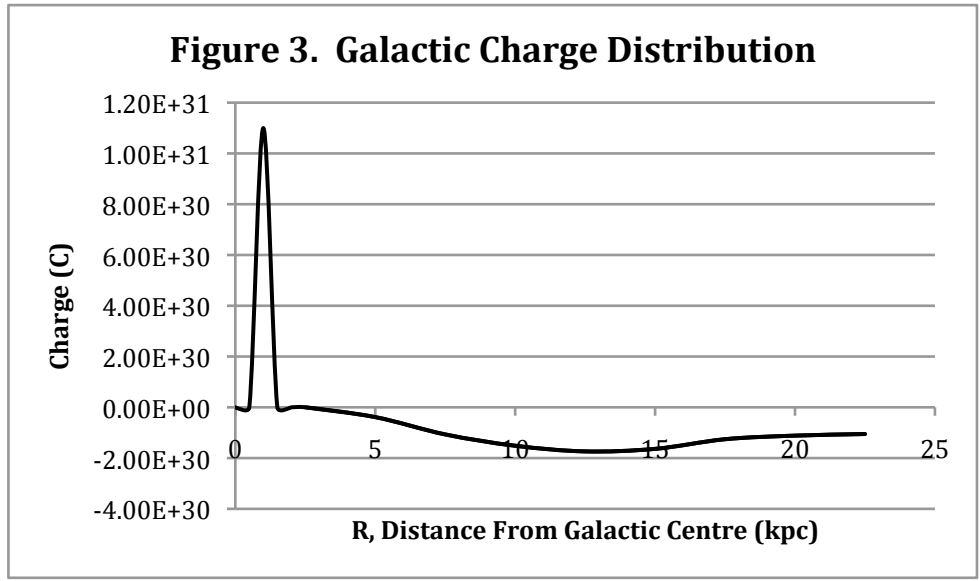

Figure 3. Galactic Charge Distribution

The final value for Q is 1.1 x $10^{31}$ C and this is shown as the initial spike in figure 3.

Our conclusion is that there is a positive charge of $\sim 10^{31}$ C on the galactic core with an equal quantity of negative charge distributed on the visible mass of the galactic periphery. This charge asymmetry gives a strong enough electric field to keep the galaxy intact with the measured rotation velocities. The detailed rotation curve behavior depends on the charge distribution assumed in the galactic periphery, but the total quantity of electrical charge needed is almost independent of the details of the charge distribution.

At first glance, this might seem like a large charge; however, it consists of less than one part in $10^{17}$ of the available charge in the galactic core. It corresponds to approximately one free proton for every $10^{11}$ m$^3$ of galactic core volume. Such a galactic core charge would produce an experimentally detectable galactic electric field at the location of the solar system of approximately 1 V/m.

It is a reasonable question to ask why there should be a charge asymmetry in the galaxy.

In fact, given the extreme and violent activity in a galactic core, it is probable that the core should evolve with a net positive electrical charge. It is quite implausible that the core should remain electrically neutral. For example, it has been estimated that there is a core-collapse supernova somewhere in the Milky Way galaxy every 50-100 years [3]. Most of these are in the galactic core. A typical supernova star contains ~ $10^{58}$ protons plus an equal number of electrons. It only takes ~ $10^{50}$ electrons moving at high velocity away from the core to leave a positive core charge of $10^{31}$ C and dump $10^{31}$ C into the galactic periphery. Therefore, the same processes that give the core a positive charge would cause the outer regions of the galaxy to develop a net negative electrical charge. It is worth noting that normal (i.e. Main Sequence) stellar processes are not expected to cause a star to develop a significant positive charge.

It is interesting that this model would interpret a galaxy as an electromagnetic dipole and, depending on their relative orientation, some of the galaxies in a galactic cluster would exhibit attractive dipole-dipole interactions in addition to gravity. Also, positively charged super novae remnants might have an anomalous motion with respect to the galactic core.

In conclusion, a simple calculation shows that we are able to reproduce the galactic rotation curve of figure 1 if we assume a net galactic core charge of approximately $10^{31}$ C with an equal and opposite charge distributed throughout the galactic periphery. A similar calculation for the Andromeda Galaxy (M31) produces a similar conclusion with galactic core and periphery charges of approximately $3 \times 10^{31}$ C.

We are not reviewing the evidence for Dark Matter in this paper, although we are aware that there are other experimental results that have been interpreted as evidence for Dark Matter. Our very simple calculations indicate that the so-called anomalous galactic rotation curves can be explained without assuming the existence of Dark Matter.

Acknowledgements, etc.

It is a pleasure to acknowledge very useful discussions with Phil Butler of the University of Canterbury, New Zealand, Robert Harrington of the University of Edinburgh, U.K. and Shomeek Mukhopadhyay of Yale University. Errors, misconceptions, etc. are of course the authors responsibility.

We are aware of several workers who have published discussions of the role of electromagnetism and plasma in the universe under the generic title, "Electric Universe" (EU). We emphasise that we are not involved with this group of researchers and we neither support nor oppose the EU ideas.